\documentstyle[prl,twocolumn,aps,epsf]{revtex}
\begin{document}
\draft

\twocolumn[\hsize\textwidth\columnwidth\hsize\csname
@twocolumnfalse\endcsname

\title{\bf    Two   band/two    gap   superconductivity   in
carbon-substituted MgB$_2$ evidenced by
point-contact spectroscopy}
\author{P. Samuely,$^{1}$ Z.  Ho\v lanov\'a,$^{1}$  P. Szab\'o,$^{1}$
J. Ka\v cmar\v c\'{\i}k,$^{1}$
R. A. Ribeiro,$^{2}$ S. L. Bud'ko,$^{2}$ P. C. Canfield,$^{2}$}
\address{$^1$Centre  of  Low   Temperature  Physics  of  the
Institute of Experimental Physics SAS
\& Faculty of Science UPJ\v S,
SK-04353~Ko\v{s}ice,         Slovakia\\
$^2$ Ames Laboratory and Department of Physics and Astronomy,
Iowa State University, Ames, IA 50011 USA}
\date{\today}
\maketitle

\begin{abstract}

The Andreev  reflection measurements of  the superconducting
energy   gap   in  the   carbon-substituted  MgB$_2$  are
presented. Despite the strong  suppression of the transition
temperature by 17 K in comparison with the pure MgB$_2$, the
same reduced  value of the small  superconducting energy gap
with  $2\Delta/kT_c  \approx$  1.7  has  been systematically
observed.   This  indicates   that  the   two  band/two  gap
superconductivity is still preserved here.

\end{abstract}

\pacs{PACS     numbers: 74.50.+r, 74.70.Ad, 74.62.Dh}

]

MgB$_2$  -  a  surprising   inter-metallic  superconductor  at  39
K \cite{akimitsu}  represents a  spectacular example  of
two band/two  gap superconductivity \cite{suhl}.  Among many
experiments         \cite{bouquet,giubileo,iavarone,suderow,eskildsen,kwok}
the point-contact spectroscopy based on the Andreev
reflection  process gave  one of  the first  proofs of  such
multi-gap superconductivity  \cite{szabo}. In line  with the
theoretical  predictions  \cite{liu},  for  the  larger  gap
$\Delta   _{\sigma}$  attributed   to  the   two-dimensional
$\sigma$-band parallel to the  $c$-axis originating from the
boron  $p_{x-y}$ orbitals,  the reduced  gap value  $2\Delta
_{\sigma} /k_BT_c \simeq 4$ has  been found. The smaller gap
$\Delta  _{\pi}$   on  the  3D   $\pi$-band  of  the   boron
$p_z$-orbitals has the reduced value much below the BCS weak
coupling  limit  of   a  one-band  superconductor  ($2\Delta
_{\pi}  /k_BT_c  \simeq  1.7  $).  One  of the fundamental
consequences   of     multigap   superconductivity   is
a breakdown   of   the   Anderson's   theorem   saying  that
superconductivity   is   not   sensitive   to   non-magnetic
impurities. Indeed, the theoretical calculations of Liu {\it et
al.} \cite{liu}  argued that the introduction  of strong defects
should have  fatal consequences for  the supeconductivitz in
MgB$_2$, merging its two
distinct  gaps  to  the  one  and  decreasing the transition
temperature  to about  22 K.  $T_c$ of  22 K  is exactly the
transition    temperature    of    the    carbon-substituted
Mg(B$_{0.9}$C$_{0.1})_2$    which    have    been   recently
synthesized  \cite{ribeiro,avdeev}. Here  we show  that despite the
strong suppression of the transition temperature by 17 K due
to the replacement of boron by carbon in comparison with the
pure  MgB$_2$  the  small  superconducting  energy  gap with
$2\Delta  /k_BT_c  \simeq  $  1.7  has  been systematically
observed closing at the bulk $T_c$ = 22 K. Its reduced value
is very similar to  the case of pure MgB$_2$ indicating
that it is $\Delta _{\pi}$ on the 3D $\pi$-band and that the
two band/two gap superconductivity is still preserved here.

Samples  of carbon-substituted  MgB$_2$ were  synthesized in
the  form of  pellets following  the procedure  described in
Ref.\cite{ribeiro} from  magnesium lumps and  B$_4$C powder.
The      nominal     stoichiometry      was     kept      as
Mg(B$_{0.8}$C$_{0.2})_2$. Synthesis temperature and time was
optimized to  1100 $^{\circ} $C and  24 hours, respectively.
Traces of B$_4$C were not visible in the XRD patterns. Small
amounts
of two    impurity  phases  (MgO  and
MgB$_2$C$_2$)  result even with  optimization  of  the
synthesis  and  may  well  indicate  that  there is a carbon
solubility limit of  $x \sim $ 0.1 for  synthesis at ambient
pressure \cite{ribeiro,avdeev}.
The  homogeneity of the sample has been  evidenced by
a narrow transition in magnetic susceptibility with an onset
of diamagnetism  at 22 K and  electrical resistance with $R$
= 0  at  21  K.  Recent  neutron  powder diffraction studies
\cite{avdeev} on
a sample  made in  the same  way but  with the  isotopically
enriched   $^{11}$B   have   revealed   a  stoichiometry  of
Mg(B$_{0.9}$C$_{0.1})_2$.

    Point-contact   measurements  have   been  performed  on
several  pieces   coming  from  two   different  batches  of
Mg(B$_{0.9}$C$_{0.1})_2$   samples  with   $T_c$  =   22  K.
A special point-contact approaching system with a negligible
thermal  expansion allows  for  temperature dependent
measurements up to 100 K.
A standard lock-in technique
at 400 Hz was used to measure the differential resistance as
a function  of applied  voltage on  the point  contacts. The
microconstrictions  were  prepared  {\it  in  situ}  by  pressing  different
metallic  (M) tips  (copper, silver,  platinum and  tungsten
formed either mechanically or by electrochemical etching) on
different parts of the  freshly  polished  surface  of  the superconductor.
The approaching  system enabled  both the  lateral and  vertical
movements of the tip by differential screw mechanism.

 Transport       of       charge       carriers       across
a normal-metal/superconductor  (N/S) interface  involves the
process of Andreev reflection. If the N/S interface consists
of a ballistic point contact with  the electronic mean free path $l$ in
the  normal metal  bigger than  the diameter  of the contact
orifice,  the  excitation  energy  $eV$  of  charge carriers
passing  the  point  contact  is  controlled  by the applied
voltage $V$.  A direct transfer of  the charge carriers with
an excitation  energy $eV < \Delta$  is forbidden because of
the   existence  of   the   energy   gap  $\Delta$   in  the
quasiparticle  spectrum of  the superconductor.  The Andreev
reflection causes the retroreflection of a hole back into the
normal  metal with  the formation  of a  Cooper pair  in the
superconductor.  At excitation  energies above  the gap  the
transfer of quasiparticles is  again allowed. This leads to
a two  times higher  conductance of  a N/S  contact at  $V <
\Delta/e$ (zero-temperature limit) for the case of ballistic
transport with  high transmission probability  of the charge
carriers $T = 1$. Surface collisions and/or mismatch of the
Fermi  velocities  in  the  point-contact forming electrodes
leads  to the  tunneling  channel  of the  carrier transport
with  $T<< 1$,  where
negligible conductance is observed inside the
gap and peak at the gap's edge. The general case
for arbitrary transmission
$T$ between  the normal tip and  the superconductor has been
treated by Blonder, Tinkham and Klapwijk (BTK)
\cite{blonder}. In any case the voltage dependence of the conductance of
a N/S contact gives direct  spectroscopic information on the
superconducting  order parameter  $\Delta$. The  conductance
data  can be  compared with  the BTK  theory using  as input
parameters  the  energy  gap  $\Delta$,  the  parameter  $z$
(measure  for the  strength  of  the interface  barrier with
transmission  coefficient  $T  =  1/(1+z^2)$  in  the normal
state),  and  a  parameter  $\Gamma$  for the quasi-particle
lifetime broadening \cite{plecenik}. In the case of the pure
MgB$_2$ for  an important contribution  of the point-contact
current parallel to the $ab$-plane the both $\sigma$ as well
as $\pi$  bands are contributing to  the conductance. It can
be  expressed   as  a  weighted  sum   of  the  partial  BTK
conductances  $\Sigma  =  \alpha  \Sigma_{\pi}  +  (1-\alpha
)\Sigma_{\sigma}$.

Figure   1  shows   typical  examples   of  the   normalized
conductance-versus-voltage spectra obtained  for the various
M-Mg(B$_{0.9}$C$_{0.1})_2$    junctions.    All    displayed
point-contact  conductances  have  been  normalized
to the
conductance  background  at higher  voltages above the
energy  gap  with  a  smooth  interpolation  inside  the gap
voltages.  After the  first soft  touch of  the tip a smooth
background  conductance  of a tunneling  character  appeared
without  any  gap  feature.  Increasing the  tip pressure
allowed  for   a  barrier  formation   and  in  some   cases
superconducting   spectral   features   appeared.   Such  an
appearence is  due to the  optimal combination  of the good quality
of the particular  grain under the tip and  the barrier. The
long  term  stability  of  contacts  was  very  bad (in
comparison with our experience  on the pure MgB$_2$) which was very
unfavorable for measurements at different temperatures. The
resulting   point-contacts    revealed   different   barrier
transparencies  from very  metallic  interface  with $z  = $
0.35  (upper  curve)  up  to  an  intermediate  case between
metallic and tunneling barrier with  $z \sim $ 0.8. Using of
different metallic  tips did not  show any influence  on the
obtained spectra showing the  superconducting energy gap. As
shown, the three lower curves  display symmetric pair of the
peaks  indicating single  but rather  small energy  gap. The
upper  curve  shows  the  highest  transparency which causes
an  increase of  the conductance  inside the  gap due to the
Andreev reflection. All these
curves could  be fitted by  the single BTK  conductance with
the  indicated resulting  parameters $\Delta$,  $\Gamma$ and
$z$.  As the  most important,  the value  of the  gap little
scattered around  1.6 meV, the value  twice smaller than the
BCS prediction for the superconductor with $T_c =$ 22 K.

Point-contact   spectroscopy   is   a   surface  sensitive
technique.  This  gives  rise  to  the  possibility that the
smaller value of the gap could be
caused by a weakening  of the superconducting state possibly
resulting   from    a   surface   proximity    effect   with
correspondingly suppressed $T_c =  2\Delta /3.52 k_B \approx
$ 11  K.  That  is  why  it  is  necessary  to establish the
particular $T_c$ of the point-contact with such a small gap.
The temperature dependence of  the point-contact spectrum of
a specific contact is
shown in Fig. 2. The spectrum shows a  pronounced increased
conductance  inside  the  superconducting  gap  due  to  the
Andreev reflection.  The barrier strength  $z =$ 0.4  yields
also   a  well  resolved   pair  of  the gap-peaks.  Such
pronounced  spectral  features   (high  conductance  due  to
Andreev reflection  and the gap-related peaks)  are a result
of  relatively  small
smearing parameter $\Gamma = $  0.32 meV which is less than
20 \% of $\Delta = 1.67$  meV. It is worth mentioning that also
the conductance intensity is well  reproduced by the BTK fit
without   any  adjustable   parameter.
The only spurious feature is a dip outside the gap at $\pm $5 mV.
Such  dips  could  originate  from  a  redistribution of the
current path  when a critical  current is reached  in a weak
link  or a  crack  nearby  the contact  \cite{samuely}. This
efect is hardly avoidable in the powder samples.

At  elevated temperatures  above 10 K the
pair of  the gap peaks merge  to the one maximum  due to the
thermal   broadening.   Nevertheless,   from   the  smoothly
decreasing Andreev reflection maximum  it is evident that the
spectrum shows the superconducting energy gap still existing
near to  the bulk critical temperature,  when at 20 K  it is
still   not  in   the  normal   state.  Together   with  the
experimental data also the  corresponding BTK fits are shown
by  opened  circles.  During  the  fit  the barrier strength
parameter $z$  and the broadening $\Gamma$  have been fixed
once determined  at 4.2 K.

The  resulting temperature dependence  of the energy  gap
$\Delta $ is  shown in Fig. 3. Big error  bars are caused by
two  effects.   First,  there  is  an   uncertainty  in  the
normalization  of the  data at  particular temperature since
the  background conductance  was changing during  the temperature
measurements. Second, it is the occurence of  the dip interfering
with the spectrum.
Nevertheless,  one  can  notice  that  a  shape  of  the  temperature
dependence  shows   deviations  from  the   BCS  prediction.
For  comparison    we  show  in  Fig.  3 also the
temperature   dependence    of   the   small    energy   gap
$\Delta_{\pi}(T)$  obtained on  the three  junctions made on
pure MgB$_2$ \cite{szabo}.
All dependences show faster
decrease  with the  temperature  than  predicted by  the BCS
theory,  but in  line with  predictions of  Liu {\it et al.}
\cite{liu}.

The shown data unequivocally prove that the small energy gap
is related  to the bulk transition  temperature. The size of
the gap as well as its temperature dependence is remarkable
similar to that of the  small energy  gap $\Delta_{\pi}$  on the
$\pi$  band of pure  MgB$_2$, just rescaled  to the
reduced transition temperature of 22 K. The absence  of the large
gap in the measured spectra  again resembles the situation in
 pure MgB$_2$  where  less than  10 percent  of the
junctions clearly  displayed the large-gap related  maximum. While the
small gap on the isotropic $\pi$-band is always contributing
to  the spectrum,  the gap  peak from  the large  gap can be
detected   only   for   important   contribution   from  the
$ab$-plane  current.  Inhomogeneities  in  the  state-of-art
samples of Mg(B$_{0.9}$C$_{0.1})_2$ cause a relatively large
broadening  $\Gamma$-parameter  which  could  hide  possible
traces of a small contribution of the large gap.

Our   spectroscopic  finding   is  in   agreement  with  the
temperature dependent specific-heat
data  by   Ribeiro  {\it  et   al.}  \cite{ribeiro}  showing
significant  thermal  excitations  above  the  small gap for
temperatures above 10 K.

Original   theoretical  estimates   of  Liu   {\it  et  al.}
\cite{liu}  showed  that  important  scattering  between the
$\sigma$ and $\pi$ bands in MgB$_2$ should lead to averaging
of the two  gaps and decrease of $T_c$ to  about 22 K. Later
theoretical   and   experimental    studies   have   revealed
difficulties in  realizing such scattering.  MgB$_2$ samples
with  very different  resistivities at  40 K  from 0.38 $\mu
\Omega$cm  \cite{canfield}  to  25  $\mu \Omega$cm \cite{tu}
with no significant change in $T_c$ has been prepared. Mazin
{\it  et  al.}\cite{mazin}   has  shown  theoretically  that
introduction of  defects like lattice  imperfections and/or
nonstoichiometricity in  MgB$_2$ leads to  a strong increase
of the  {\it intraband} scattering  particularly inside the
$\pi$-band rather than to an {\it interband } scattering. It
is  only  the  latter  one  which  should  reveal a proposed
pair-breaking effect leading  to a one-gap superconductivity
with  depressed  $T_c$.  The  theoretical  calculations also
showed that due to a very
different $k$-space  distribution of the  $\pi$ and $\sigma$
bands,  the  only  route  to  increase  the  $\sigma  - \pi$
scattering is  via interlayer hopping, from  a $p_z$ orbital
($\pi$-band)  in   one  atomic  layer  to   a  bond  orbital
($\sigma$-band) in another layer.

Substitution     of     boron     by     carbon    in    the
Mg(B$_{0.9}$C$_{0.1})_2$ samples naturally  leads to a heavy
increase of the resistivity. Rough estimate with no account
for the porosity indicates hundreds of $\mu \Omega$cm at low
temperatures \cite{ribeiro}. The neutron diffraction studies
proved  no  ordering  of   carbon.  This  suggests  a  large
electronic   scattering  in   the  system.   In  our  recent
experiments  \cite{holan} the  upper critical  field at  1.5
K of about 30 T was found much bigger than in the pure MgB$_2$. It
indicates that the doped samples are in  a dirty limit with a very
short  mean  free  path.  On  the  other  hand  the both XRD
\cite{ribeiro}   and   neutron   experiments   \cite{avdeev}
revealed no  change in the  $c$-lattice parameter in  carbon
substituted  samples in  comparison with  the pure  MgB$_2$.
Then, conditions for interlayer hopping which could increase
the $\sigma - \pi$ scattering are not more favorable than in
MgB$_2$.  The  significant  suppression  of  $T_c$ could
probably be related to the decrease density of states and Debye
temperature in the system \cite{ribeiro} and  not due to suppression of the
two-gap superconductivity. A  significantly lower anisotropy
in  $H_{c2}$  \cite{ribeiro}  in  the  carbon doped compound
implies that  the $\sigma$-band Fermi surface  is not nearly
so 2D  as in the  pure MgB$_2$. This  can partially suppress
the  strong  electron-phonon  coupling  in the $\sigma$-band
responsible  for $T_c$.  Changes  in  the Fermi  surface are
resulting from  different electronic configuration  in boron
and carbon.

In conclusion, we have obtained an experimental evidence for
the existence  of the small  superconducting energy gap  in
the carbon-substituted MgB$_2$ closing  at the bulk $T_c$. The
regular observation  of this effect  in our spectra  and the
support  for  it   by  other  measurements  \footnote{During
completion of the paper we  noted a preprint of Schmidt {\it
et   al.}   \cite{schmidt}    with   similar   conclusions.}
demonstrates a survival of  the two-gap superconductivity in
the  carbon-substituted  samples   with  heavily  suppressed
$T_c$.


This  work  has  been  supported  by  the Slovak Science and
Technology     Assistance      Agency     under     contract
No.APVT-51-020102.  Centre  of  Low  Temperature  Physics is
operated as  the Centre of Excellence  of the Slovak Academy
of Sciences under contract no. I/2/2003.  Ames Laboratory is
operated for the U.S. Department of Energy by Iowa State University under
Contract No. W-7405-Eng-82.  This work was supported by the Director for
Energy Research, Office of Basic Energy Sciences. The liquid nitrogen for
the experiment has been sponsored  by the U.S. Steel Ko\v sice,
s.r.o.

\newpage

\onecolumn

{\bf FIGURE CAPTIONS}

\begin{figure}
\epsfverbosetrue
\epsfxsize=10cm
\epsfysize=12cm
\begin{center}
\hspace{550mm}
\epsffile{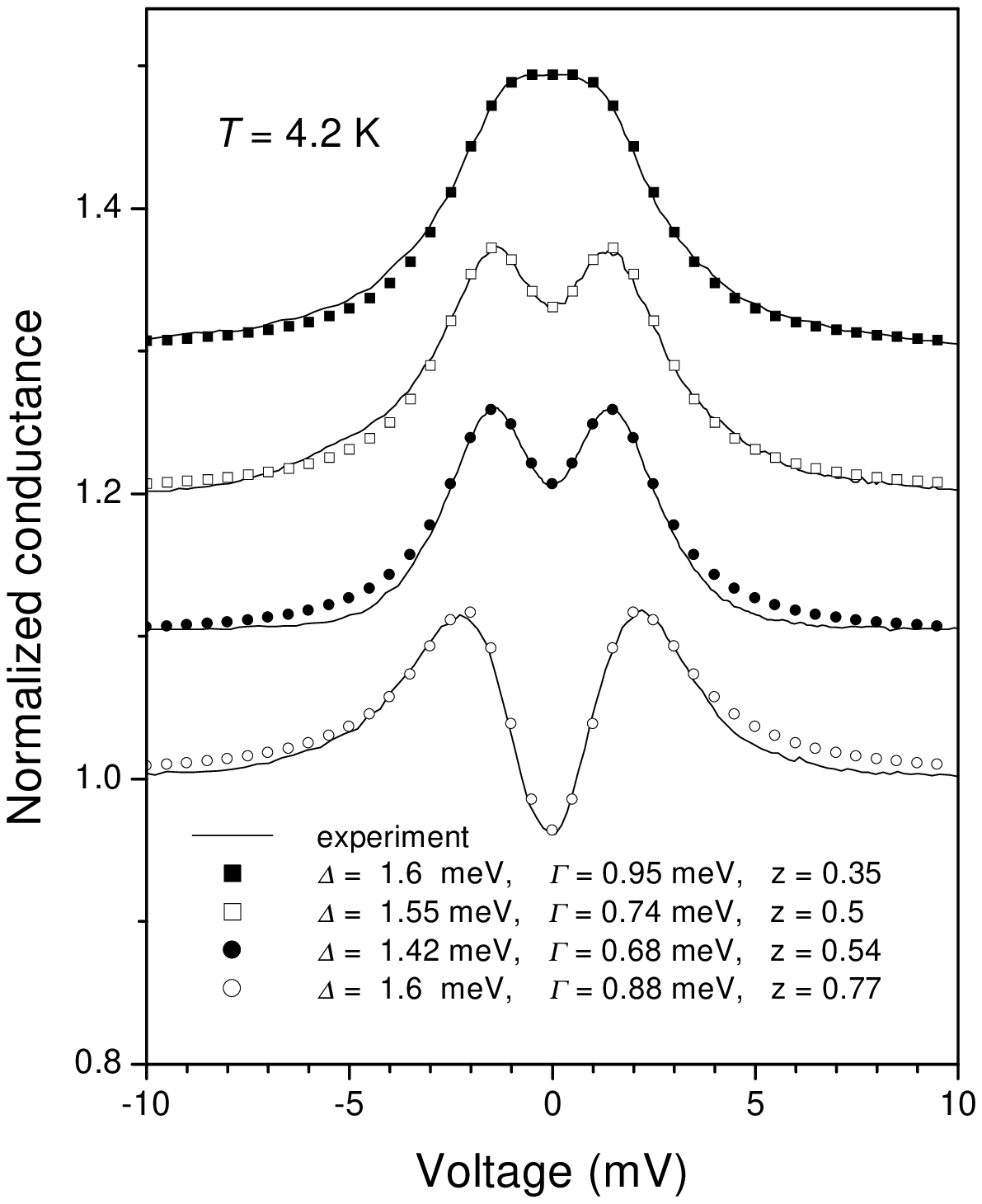}
\vspace{2mm}
\end{center}
\caption{
Metal-Mg(B$_{0.9}$C$_{0.1})_2$  point-contact spectra  at $T
=$  4.2  K  (full  lines).  The  upper curves are vertically
shifted for  the clarity. Symbols -  fitting
for the thermally smeared BTK model.}
\end{figure}

\newpage

\begin{figure}
\epsfverbosetrue
\epsfxsize=10cm
\epsfysize=12cm
\begin{center}
\hspace{550mm}
\epsffile{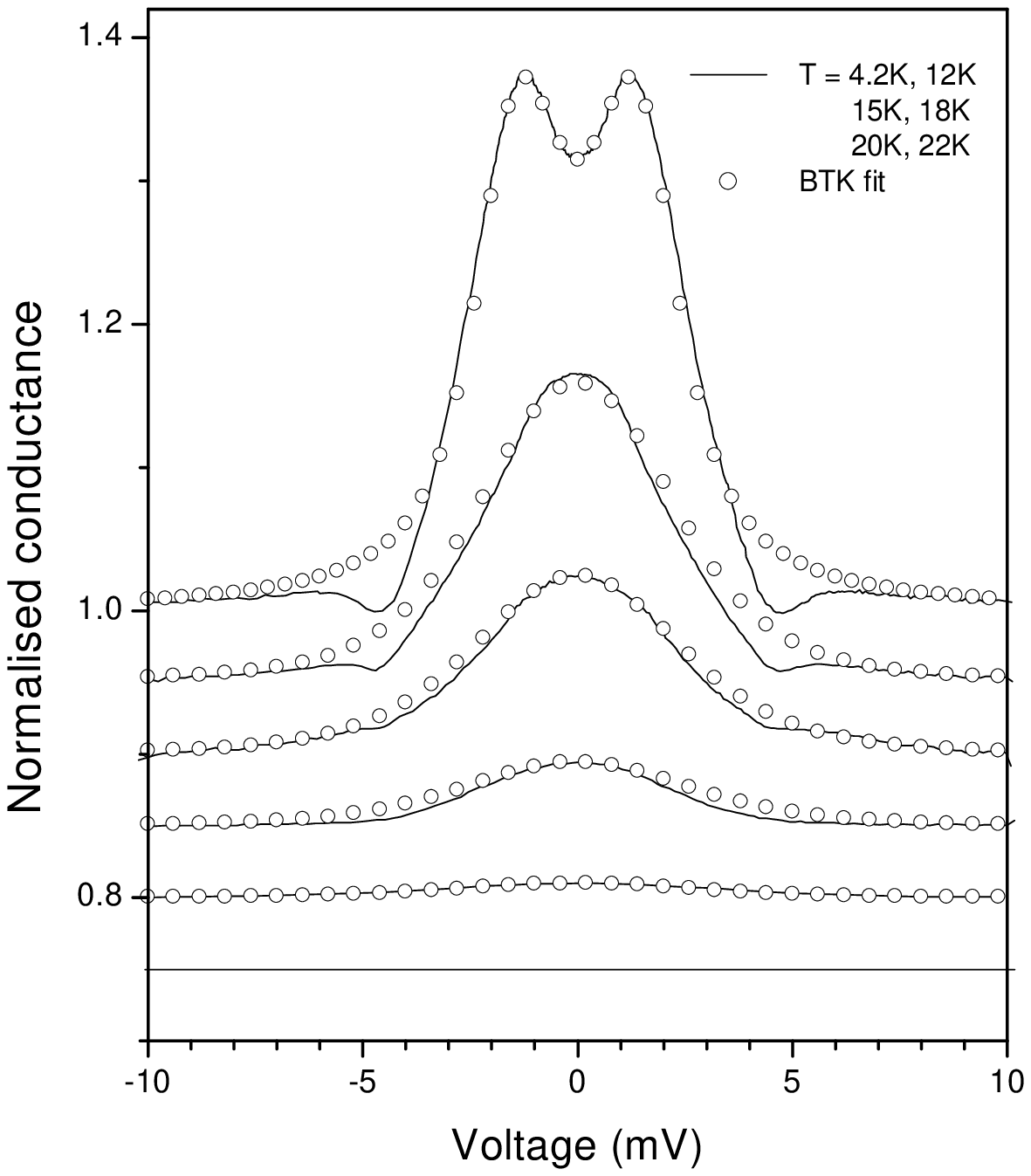}
\vspace{2mm}
\end{center}
\caption{
 Differential  conductances  of   Cu-Mg(B$_{0.9}$C$_{0.1})_2$
point-contact  measured   (full  lines)  and   fitted  (open
circles) for the thermally smeared BTK model at indicated
temperatures. The  fitting parameters $z =
0.4$, $\Gamma = 0.32$ meV and $\Delta$ (4.2 K) = 1.67 meV.
The  lower curves are vertically
shifted for  the clarity.}
\end{figure}

\newpage

\begin{figure}
\epsfverbosetrue
\epsfxsize=10cm
\epsfysize=7cm
\begin{center}
\hspace{550mm}
\epsffile{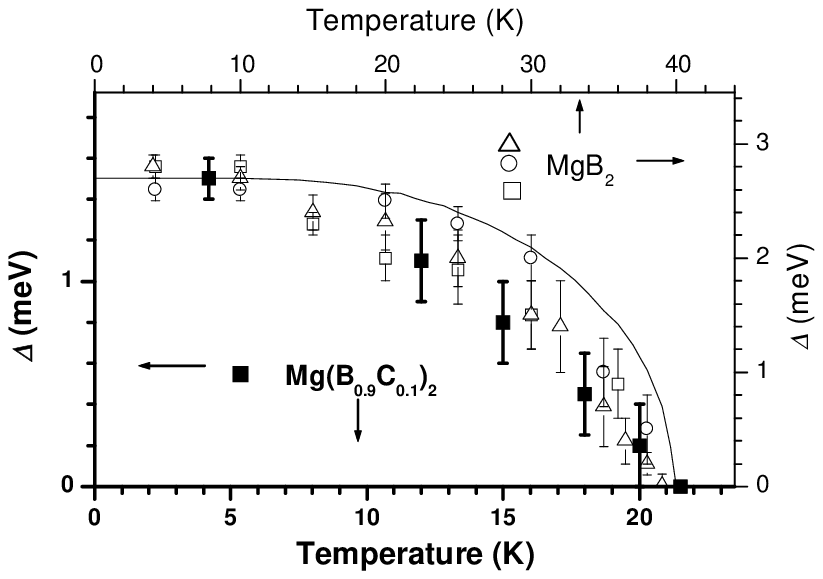}
\vspace{2mm}
\end{center}
\caption{Bold symbols - temperature   dependence   of    the   energy   gap
in Mg(B$_{0.9}$C$_{0.1})_2$ determined
from the fitting of the point-contact spectrum shown in Fig.
2. Opened symbols - temperature  dependence of the small gap
$\Delta_{\pi }$ in
the  undoped MgB$_2$  obtaimed  from fitting to
three different contacts [9].
 Full lines represent the BCS prediction.}
\end{figure}

\end{document}